\documentstyle[prl,aps,preprint]{revtex}
\begin{document}
\preprint{}

\title{Magnetoresistance in ordered and disordered
double perovskite oxide, Sr$_2$FeMoO$_6$}
\author{D. D. Sarma$^a$, E.V. Sampathkumaran$^b$, Sugata Ray$^a$, R. Nagarajan$^b$, 
Subham Majumdar$^b$, Ashwani Kumar$^a$, G. Nalini$^a$ and T.N. Guru Row$^a$}
\address{$^a$Solid State and Structural Chemistry Unit, Indian Institute
of Science, Bangalore 560 012, India}
\address{$^b$Tata Institute of Fundamental Research, Mumbai - 400 005, India}
\date{\today}
\maketitle

\begin{abstract}

We have prepared crystallographically ordered and disorder specimens of
the double perovskite, Sr$_2$FeMoO$_6$ and investigated their magnetoresistance
behaviour. The extent of ordering between the Fe and Mo sites
in the two samples is determined by Rietveld analysis of powder x-ray 
diffraction patterns and reconfirmed by M\"{o}ssbauer studies. While the 
ordered sample exhibits the sharp low-field response, followed by 
moderate changes in the magnetoresistance at higher fields, the disordered 
sample is characterised by the absence of the spectacular low-field response.
We argue that the low field response depends crucially on the 
half-metallic ferromagnetism, while the high-field response follows from the
overall magnetic nature of the sample, even in absence of the half-metallic
state.

\end{abstract}



\section{Introduction}

    Colossal magnetoresistance (CMR) is a property that is of great
technological potential, since the large change in the resistance (R)
with the application of a magnetic field can be used 
effectively for mass storage magnetic devices. However, the
well-known CMR materials, Mn based oxides, have significant CMR
effect only at low temperatures and, therefore, are not
suitable for room temperature applications.
In their recent work, Kobayashi {\it et al.}~\cite{koba} have
pointed out that the fully-ordered double perovskite
Sr$_2$FeMoO$_6$ with alternating Fe$^{3+}$ (3$d^5$, $S=5/2$) and
Mo$^{5+}$ (4$d^1$, $S=1/2$) ferrimagnetically coupled ions exhibit
substantial CMR even at room temperature. It is suggested that the
half-metallic ferromagnetic (HMFM) state below T$_c$ is responsible
for the magnetoresistance (MR) behavior via spin-dependent carrier
scattering processes. 
While the spin-dependent scattering due to the
intergrain tunneling effects enables one to construct manganite
based devices~\cite{mathur}, the intra-grain properties of the
manganites are generally believed to be intimately connected with
the intrinsic instability of Mn$^{3+}$ 3$d^4$ states arising from
strong electron-phonon interaction as well as double-exchange
interaction~\cite{millis1,millis2}. 
Mo$^{5+}$ is also a Jahn-Teller ion with its 4$d^1$ $^2D$
degenerate ground state. Thus, {\it a priori} it is not possible to
rule out any contribution to the magnetoresistance of
Sr$_2$FeMoO$_6$ from effects other than intergrain spin-dependent carrier
scatterings. 
Considering the
complexity of the mechanism of magnetoresistance in the manganites,
in which in addition to half-metallic ferromagnetism, various
instabilities play important role, it is essential to address the
factors responsible for the novel magnetoresistance properties of
Sr$_2$FeMoO$_6$.

Central to the HMFM state in Sr$_2$FeMoO$_6$ as deduced by the band
structure calculation~\cite{koba} is the long-range ordering of
FeO$_6$ and MoO$_6$ octahedra alternatingly along three cubic axes.
Disorder is
expected to be detrimental to the half-metallic nature, giving rise
to finite up- and down-spin densities of states at the Fermi
energy, destroying the HMFM state; preliminary super-cell band
structure calculations in our group indeed suggest that HMFM state
is easily destroyed by a positional disorder at the Fe/Mo sites,
though an overall ferrimagnetic order persists in the system.
Keeping this in mind we have prepared for the first time
Sr$_2$FeMoO$_6$ by melt-quenching, where Fe and Mo sites are
heavily disordered. A comparative study of disordered and ordered
samples using x-ray diffraction, $^{57}$Fe M\"{o}ssbauer and 
the magnetoresistive properties 
allows us to delineate the effects
arising from long range crystallographic order and those from
factors other than half-metallic ferromagnetism in controlling the
MR properties. 

\section{Experimental}

We prepared ordered Sr$_2$FeMoO$_6$ following ref.~1 by hydrogen reduction of a 
mixture of SrCO$_3$, MoO$_3$ and Fe$_2$O$_3$. The disordered sample was prepared
by melt-quenching of a mixture of SrCO$_3$, MoO$_3$, Fe$_2$O$_3$ and Mo-metal
in an argon atmosphere. M\"{o}ssbauer study was carried out in 
transmission geometry using a $^{57}$Co/Rh matrix source. High resolution 
powder diffraction data were collected on a STOE STADI/P diffractometer. The 
resistivity measurements with and without an applied field were carried 
out using the standard four-probe dc measurements.

\section{Results and discussion}

We show the powder x-ray diffraction patterns of the two samples in Fig. 1;
Fig. 1(a) shows the pattern of the sample prepared by the conventional solid
state techniques \cite{koba}, while Fig. 1(b) shows the same for the 
melt-quenched sample prepared by the arc melting. The main diffraction peaks
are the same in both the panels of Fig. 1 and these are readily ascribed to 
the perovskite lattice. The only difference between the two diffraction 
patterns is in terms of the existence of two weak peaks, at 19.6$^\circ$ 
and 38$^\circ$ in Fig. 1(a), but absence in Fig. 1(b). We have expanded
the vicinity of these two angles in order to illustrate this point clearly
in the respective panels. The existence of these two peaks in Fig. 1(a) 
establishes the presence of the supercell arising from the ordering of
Fe and Mo sites alternatingly, giving rise to the double perovskite 
structure. Total or near-absence of these two order-related peaks in Fig. 1(b)
suggests that the melt-quenched sample obtained from the arc furnace 
gives rise to heavy disrodering between the Fe and the Mo sites
The extent of ordering at the Fe/Mo sites is easily estimated from
x-ray diffraction (XRD) of ordered ({\bf A}) [Fig.~1(a)] and
disordered ({\bf
B}) [Fig.~1(b)] samples of Sr$_2$FeMoO$_6$. Rietveld refinement for both 
converged to the same cell dimensions
($a$~=~$b$~=~5.566~{\AA} and $c$~=~7.858~{\AA}) with
R$_{wp}\le$~8.0\%. The ordering at the Fe and Mo sites in sample {\bf A}   
was found to be
$\sim$~91\%, slightly larger than the
previously reported~\cite{koba} value of 87\%. 
Near-absence of the long-range order-related peaks within the
noise level of the data [see insets I and II to Fig.~1(b)] for
sample {\bf B} confirms the extensive disorder in this case.
Rietveld analysis suggests approximately 31\% ordering in this case.

The effect of disorder at the Fe-sites is also probed by $^{57}$Fe
M\"{o}ssbauer spectra of these two samples at 300 K and 4.2~K (Fig.~2);
a comparison of the raw spectra for the two samples at these
temperatures is sufficient to establish the relative degrees of
disorder in site occupancies. The spectra for the sample {\bf B}
exhibits relatively larger line-widths in contrast to those of sample
{\bf A}, implying a larger disorder in sample {\bf B}. All the
spectra are magnetic hyperfine split, indicating magnetic ordering
of the materials. The spectrum at 300 K for sample {\bf A}
favorably compares with that reported earlier~\cite{nak}. Our
analysis of the data at 4.2 K for sample {\bf A} yields a
saturation hyperfine field of about 480 kOe, with a small
distribution of about 10 kOe. This value is somewhat smaller than
that known for Fe$_2$O$_3$. The hyperfine field for sample {\bf B}
has a much larger distribution (490 $\pm$~60 kOe), quantitatively
confirming a strong variation in the chemical environment for Fe
from site to site. 
The chemical composition was however found to
be identical between the two samples by energy
dispersive x-ray analysis. Moreover, the grain size and morphology
were also found to be similar between the two samples by scanning
electron microscopy (SEM). The above results establish that it is
indeed possible to obtain highly ordered and extensively disordered
samples with respect to Fe and Mo occupancies without otherwise
disturbing the structural integrity of this double perovskite
system.

We show the magnetoresistance, $MR(H,T)=\{R(H,T)-R(0,T)\}/R(0,T)$,
of ordered and disordered samples as a function of applied magnetic
field, H, at 300~K and 4.2~K in Fig~3; in the same figure we
reproduce the results from Kobayashi {\it et al.}~\cite{koba}. 
At both the temperatures, the ordered sample is characterised by 
a sharp magnetoresistive response in the low-field region, though the 
magnitude of the magnetoresistance is considerably larger
at the lower temperature, in agreement
with the published literature \cite{koba}.
The present ordered sample has
a sharper low-field response and
a larger magnitude, particularly at the low temperature compared
to the previous result~\cite{koba}. This improved
magnetoresistance response is possibly related to somewhat
higher degree of ordering in the present sample. 
This improvement in the low magnetic field response
suggests that a fully (100\%) ordered sample is
likely to be a very good candidate for real-life device
applications. 
Beyond about 1~T, the magnetoresistance of the ordered sample
exhibits a slower change at higher field strengths, without 
showing any signs of saturation upto the highest magnetic field
probed in the present experiments.
The MR changes significantly (by about 6.5\% at 4.2~K
and 3\% at 300~K) in the larger field regime between 1 and 7~T.

In contrast, the MR of the disordered sample does not exhibit
the low-field sharp magnetic response below 1~T.
Since the two samples differ only in the extent of
Fe/Mo ordering, 
the low-field rapid variation in MR
of sample {\bf A} evidently arises from the long-range order,
leading to the half metallic ferromagnetic 
state of sample {\bf A} and the consequent strong intergrain
spin-dependent scattering. It is, however, intriguing to note that 
there is a substantial negative magnetoresistance in the disordered sample
and the high
field behaviors of the MR in the  two samples are similar,
the two curves being approximately parallel in the high-field
region at both the temperatures shown in Fig.~3. This implies that
there is a common intragrain origin of this high-field MR behavior
in the disordered and the ordered samples. 

We have carried out band structure calculations, simulating
the disordered system within a supercell approach where the Fe and Mo atoms
were allowed to occupy the other sites in contrast to the perfectly ordered 
system.  Preliminary results from these calculations show that 
in the disordered state, both up- and down-spin states exist at 
the Fermi energy; this indicates that the 
half-metallic ferromagnetic ground state is not realised for any disordering
of the Fe and the Mo sublattices, though the system remains to be a ferrimagnet.
Thus, the high-field, intragrain contribution to the negative 
magnetoresistance is 
independent of the long range ordering of Fe and Mo atoms and 
the cosnequent existence of
the HMFM state. Instead, it is to be associated with the usual negative 
magnetoresistance observed for ferromagnetic substances, arising from a 
suppression of the spin-fluctuations with the application of an external
magnetic field.

In conclusion, we have prepared highly ordered and disordered samples of 
the perovskite oxide, Sr$_2$FeMoO$_6$. We characterised the extent of 
ordering by x-ray diffraction and M\"{o}ssbauer studies of these samples.
Rietveld analysis of the powder diffraction pattern shows that the extent 
of ordering in the Fe/Mo sites is about 91\% for the ordered sample, 
while it is about 31\% for the disordered sample. We point out that the 
disordered sample is not a half-metallic ferromagnet, though it is magnetic
even at the room temperature. We showed that the magnetoresistance of the 
two samples are distinguished by the existence of the sharp changes in the 
magnetoresistance at low fields for the ordered sample. In the case of the
disordered sample, there is no such sharp changes in the MR, though a negative
MR is seen at all fields for the disordered sample also. The MR of the 
disordered sample is very similar to that of the ordered sample in the 
high-field regime, indicating a common origin. Thus, we conclude that the
low-field magnetoresistance is dominated by the intergrain spin-dependent 
scattering of the highly spin polarized charge-carriers of the half-metallic
ferromagnet, the high-field negative magnetoresistance behaviours in both 
the samples are intrinsic, intragrain property and 
arise from the suppression of spin-fluctuations under the 
application of a magnetic field on a magnetic system.

{\bf Acknowledgements:} The authors thank Prof. B. Sriram Shastry
and Dr. Tanusri Saha Dasgupta for useful discussions.

\newpage

\section{figure captions}
Fig.~1. X-ray diffraction patterns of (a) ordered 
and  (b) disordered Sr$_2$FeMoO$_6$. 
Order-related peaks, appear at 19.6$^\circ$ and 38$^\circ$,
shown
in the insets on an expanded scale. 

Fig.~2.  
$^{57}$Fe M\"{o}ssbauer spectra of the disordered and ordered Sr$_2$FeMoO$_6$
 at 300 and 4.2 K. 

Fig.~3. 
The comparisons of percentage magnetoresistance between
the ordered and disordered  samples as well as the
result presented by Kobayashi {\it et al.}~\cite{koba} at (a) 4.2 K
 and (b) 300 K.

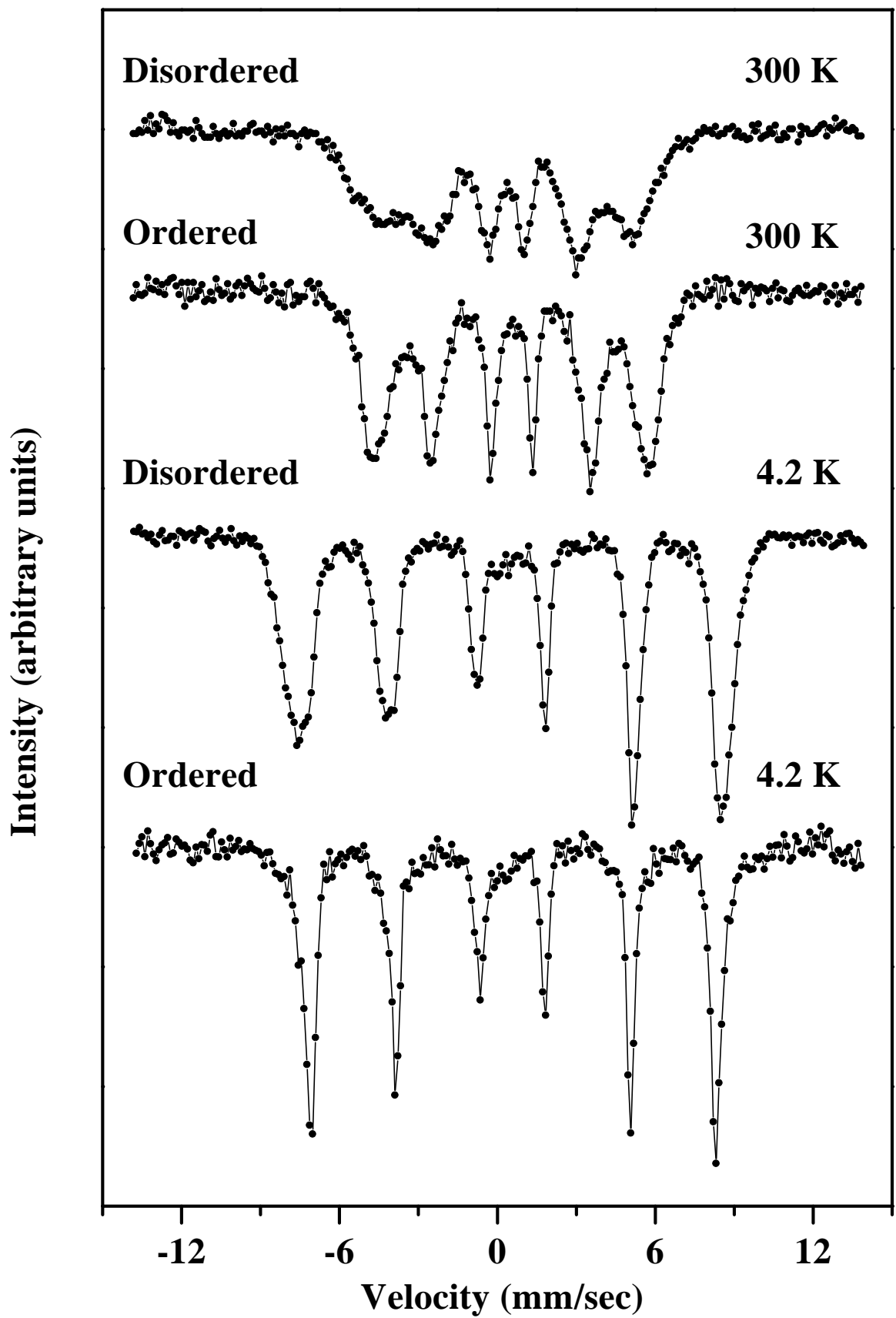
\end{document}